\documentstyle[12pt]{article}
\input epsf.tex

\newcommand{\nc}{\newcommand}
\nc{\beq}{\begin{equation}}
\nc{\eeq}{\end{equation}}
\nc{\beqa}{\begin{eqnarray}}
\nc{\eeqa}{\end{eqnarray}}
\nc{\lra}{\leftrightarrow}
\def\sfrac#1#2{{\textstyle{#1\over #2}}}
\nc{\sss}{\scriptscriptstyle}
{\nc{\lsim}{\mbox{\raisebox{-.6ex}{~$\stackrel{<}{\sim}$~}}}
{\nc{\gsim}{\mbox{\raisebox{-.6ex}{~$\stackrel{>}{\sim}$~}}}

\def\eps{\epsilon}

\begin{document}

\begin{titlepage}
\begin{flushright}
McGill 00-25\\
hep-ph/0008185\\
\end{flushright}

\vskip.5cm
\begin{center}
{\large{\bf 5-Dimensional Warped Cosmological Solutions
 \vskip0.4cm
With Radius Stabilization by a Bulk Scalar}}
\end{center}
\vskip1.5cm

\centerline{ James M.\ Cline and Hassan Firouzjahi}

\centerline{Physics Department, McGill University,
Montr\'eal, Qu\'ebec, Canada H3A 2T8}

\begin{abstract}
\vskip 3pt

We present the 5-dimensional cosmological solutions in the Randall-Sundrum
warped compactification scenario, using the Goldberger-Wise mechanism to
stabilize the size of the extra dimension.  Matter on the Planck and TeV
branes is treated perturbatively, to first order.  The back-reaction of the
scalar field on the metric is taken into account. We identify the
appropriate gauge-invariant degrees of freedom, and show that the
perturbations in the bulk scalar can be gauged away.  We confirm previous,
less exact computations of the shift in the radius of the extra dimension
induced by matter.  We point out that the physical mass scales on the TeV
brane may have changed significantly since the electroweak epoch due to
cosmological expansion, independently of the details of radius stabilization.

\end{abstract}

\vfill
\end{titlepage}

{\bf 1. Introduction.}
The Randall-Sundrum (RS) idea \cite{RS} for explaining the weak-scale
hierarchy problem has garnered much attention from both the phenomenology
and string-theory communities, providing a link between the two which is
often absent.  RS is a simple and elegant way of generating the TeV scale
which characterizes the standard model from a set of fundamental scales
which are of order the Planck mass ($M_p$).  All that is needed is that the
distance between a hidden and a visible sector brane be approximately $b =
35/M_{p}$ in a compact extra dimension, $y\in[0,1]$.  The warping of space
in this extra dimension, by a factor $e^{-kby}$, translates the moderately
large interbrane separation into the large hierarchy needed to explain the
ratio TeV$/M_{p}$. 

However the RS idea as originally proposed was incomplete due to the
lack of any mechanism for stabilizing the brane separation, $b$.  This
was a modulus, corresponding to a massless particle, the radion, which
would be ruled out because of its modification of gravity: the
attractive force mediated by the radion would effectively increase
Newton's constant at large distance scales.  An attractive model for
giving the radion a potential energy was proposed by Goldberger and
Wise (GW) \cite{GW2}; they introduced a bulk scalar field with different
VEV's, $v_0$ and $v_1$, on the two branes.  If the mass $m$ of the
scalar is small compared to the scale $k$ which appears in the warp
factor $e^{-kby}$, then it is possible to obtain the desired interbrane
separation.  One finds the relation $e^{-kb} \cong (v_1/v_0)^{4k^2/m^2}$.

An important benefit of stabilizing the radion is that cosmology is
governed by the usual Friedmann equations, up to small corrections of order
$\rho/$(TeV)$^4$ \cite{CGRT}.  Even with stabilization, there may be a
problem with reaching a false minimum of the GW radion potential
\cite{CF1}, but without stabilization, there is a worse problem: an
unnatural tuning of the energy densities on the two branes is required for
getting solutions where the extra dimension is static \cite{CGKT, CGS}, a
result which can be derived using the (5,5) component of the Einstein
equation $G_{mn} = \kappa^2 T_{mn}$.  However when there is a nontrivial
potential for the radius, $V(b)$, the (5,5) equation serves only to
determine the shift $\delta b$ in the radius due to the expansion, and
there is no longer any constraint on the matter on the branes.  Although
this point is now well appreciated \cite{KKOP}-\cite{Kim}, it has not
previously been explicitly demonstrated by solving the full 5-dimensional
field equations using a concrete stabilization mechanism.  Indeed, it has
been claimed recently that such solutions are not possible with an
arbitrary equation of state for the matter on the branes
\cite{KP}-\cite{Enq}, and also that the rate of expansion does not
reproduce normal cosmology on the negative tension brane despite
stabilization \cite{MB}. Our purpose is to  present the complete solutions,
to leading order in an expansion in the energy densities on the branes,
thus refuting these claims.

{\bf 2. Preliminaries.} 
The action for 5-D gravity coupled to the stabilizing scalar field 
$\Phi$ and matter on the branes (located at $y=0$ and $y=1$,
respectively) is 
\beqa
S &=& \int d^{\,5}x\sqrt{g} \left( -{1\over 2\kappa^2}R - \Lambda +
    \sfrac{1}{2}\partial_{\mu}\Phi\partial^{\mu}\Phi
	-V(\Phi)\right)\nonumber\\
	&+&\int d^{\,4}x \sqrt{g}\left({\cal L}_{m,0} -
V_{0}(\Phi)\right)|_{y=0}
	+\int d^{\,4}x \sqrt{g}\left({\cal L}_{m,1} -
	V_{1}(\Phi)\right)|_{y=1},
\eeqa
where $\kappa^2$ is related to the 5-D Planck scale $M$ by $\kappa^2 = 
1/(M^3)$.  The negative bulk cosmological constant needed for the RS
solution is parametrized as $\Lambda = -6 k^2/\kappa^2$
and the scalar field potential is that of a free field,
$V(\Phi) = \sfrac12 m^2 \Phi^2$.  The brane potentials $V_0$ and $V_1$
can have any form that will insure nontrivial VEV's for the scalar field
at the branes, for example $V_i(\Phi) = \lambda_i(\Phi^2-v_i^2)^2$ \cite{GW2}.
In ref.\ \cite{CF1} we pointed out that the choice $V_i(\Phi) =m_i(\Phi -
v_i)^2$ is advantageous from the point of view of analytic calculability
(see also \cite{DeWolfe}).

We will take the metric to have the form
\beqa
	ds^{2} &=& n^2(t,y)dt^{2} 
        -a^2(t,y)\sum_i dx_i^2 -b^{2}(t,y)dy^{2}
	\nonumber \\
               &=& e^{-2N(t,y)}dt^2-
		a_0(t)^{2}e^{-2A(t,y)}
		\sum_i dx_i^2 -b(t,y)^{2}dy^{2},
\eeqa
where a perturbative expansion in the energy densities of the branes
will be made around the static solution:
\beqa
		N(t,y) &=& A_0(y) + \delta N(t,y); \qquad
		A(t,y) = A_0(y) + \delta A(t,y) \nonumber\\
		b(t,y) &=& b_0 + \delta b(t,y); \qquad\qquad
		\Phi(t,y) = \Phi_0(y) + \delta\Phi(t,y).
\eeqa
The perturbations are taken to be linear in the energy densities $\rho_*$
and $\rho$ of matter on the Planck and TeV branes, located at
$y=0$ and $y=1$, respectively.

This ansatz is to be substituted into the Einstein equations, $G_{mn} =
\kappa^2 T_{mn}$, and the scalar field equation
\begin{equation}
	\partial _{t}\left(\frac{1}{n}ba^{3}\dot{\Phi}\right)
	-\partial_{y}\left(\frac{1}{b}a^{3}n\Phi^{\prime}\right)
	+ba^{3}n\left[ V'
	+ V_{0}'\delta(by)
	+ V_{1}'\delta(b(y-1)) \right] =0.
\end{equation}
Here and in the following, primes on functions of $y$ denote
${\partial\over \partial y}$, while primes on potentials of $\Phi$
will mean ${\partial\over \partial \Phi}$.
The nonvanishing components of the Einstein tensor are
\beqa
G_{00} &=& 3\left[(\frac{\dot {a}}{a})^{2}+\frac{\dot {a}}{a}
	\frac{\dot {b}}{b}-\frac{n^{2}}{b^{2}}\left(\frac {a^{\prime \prime}}{a}
	+(\frac{a^{\prime}}{a})^{2}
	-\frac{a^{\prime}b^{\prime}}{ab}\right)\right]\nonumber\\
G_{ii}&=&\frac{a^{2}}{b^{2}}
	\left[(\frac{a^{\prime}}{a})^{2}
	+2\frac{a^{\prime}}{a}\frac{n^{\prime}}{n}
	-\frac{b^{\prime}}{b}\frac{n^{\prime}}{n}-	
	2\frac{b^{\prime}}{b}
	\frac{a^{\prime}}{a}+2\frac{a^{\prime
	\prime}}{a}+\frac{n^{\prime \prime}}{n}\right]\nonumber\\
	&+&\frac{a^{2}}{n^{2}}\left[-(\frac{\dot
	{a}}{a})^{2}+2\frac{\dot{a}}{a}\frac{\dot{n}}{n}
	-2\frac{\ddot{a}}{a}+\frac{\dot{b}}{b}(-2\frac{\dot{a}}{a}
	+\frac{\dot{n}}{n})-\frac{\ddot {b}}{b}\right]\nonumber\\
G_{05}&=&3\left[\frac{n^{\prime}}{n}\frac{\dot{a}}{a}
	+\frac{a^{\prime}}{a}\frac{\dot{b}}{b}
	-\frac{\dot{a}^{\prime}}{a}\right]\nonumber\\
G_{55}&=&3\left[\frac{a^{\prime}}{a}\left(\frac{a^{\prime}}{a}
	+\frac{n^{\prime}}{n}\right)-\frac{b^{2}}{n^{2}}
	\left(\frac{\dot{a}}{a}\left(\frac{\dot{a}}{a}
	-\frac{\dot{n}}{n}\right)+\frac{\ddot{a}}{a}\right)\right]
\eeqa
and the stress energy tensor is
$T_{mn} = g_{mn}(V(\Phi)+\Lambda)+\partial_{m}\Phi\partial_{n}\Phi
-\frac{1}{2}\partial^{l}\Phi\partial_{l}\Phi g_{mn}$ in the bulk.
On the branes, $T_m^n$ is given by
\begin{eqnarray}
T_{m}^{n} &=&\delta(by)\,{\rm 
diag}(V_{0}+\rho_{*},V_{0}-p_{*},V_{0}-p_{*},V_{0}-p_{*},0)\nonumber
\\
&+&\delta(b(y-1))\,{\rm diag}(V_{1}+\rho,V_{1}-p,V_{1}-p,V_{1}-p,0)
\end{eqnarray}

At zeroth order in the perturbations, the equations of motion can be
written as
\beqa
	{A_0^{\prime}}^{2} &=& \frac{\kappa^{2}}{12}
 	\left({\Phi_0^{\prime}}^{2}-
	m^2b_{0}^{2}\Phi_0^{2}\right) +k^{2}b_{0}^{2}; \qquad
	A_0^{\prime \prime}=\frac{1}{3}\kappa^{2}{\Phi_0^{\prime}}^{2}
	\nonumber\\
\label{zeroeq}
	\Phi_0^{\prime \prime}&=&4A_0^{\prime}
	\Phi_0^{\prime}+m^2b_{0}^{2}\Phi_0,
\eeqa
and the solutions are approximately 
\beq
\label{approxsoln}
	\Phi_0(y) \cong v_0 e^{-\eps k b_0 y};\qquad
	A_0(y) \cong kb_0y + \frac{\kappa^2}{12}v_0^2 (e^{-2\eps kb_0y} -1)
\eeq
where we have normalized $A_0(0) = 0$, and introduced
\beq
	\eps = \sqrt{4 + {m^2\over k^2}} - 2  \cong {m^2\over 4 k^2}.
\eeq
The above approximation is good in the limit $\eps\ll 1$, which is
the same regime in which the Goldberger-Wise mechanism naturally gives
a large hierarchy without fine-tuning the scalar field VEV's on the branes:
$e^{-kb_0} = (v_1/v_0)^{1/\eps}$.  For small $\eps$, the GW solution
coincides
with an exact solution of the coupled equations that was presented in
ref.\ \cite{DeWolfe}.

{\bf 3. Perturbation Equations.}  We can now write the equations for the
perturbations of the metric, $\delta A$, $\delta N$, $\delta b$,
and the scalar field, $\delta \Phi$.  The equations take a simpler form
when expressed in terms of the following combinations:
\beq
	\Psi = \delta A' - A'_0 {\delta b\over b_0}-\frac{\kappa^2}{3} \Phi_0' \delta\Phi;\qquad
	\Upsilon = \delta N' - \delta A'
\eeq
Further simplification comes from realizing that the perturbations
will have the form, for example, $\Psi = \rho_*(t) g_0(y) +
\rho(t) g_1(y)$, so that their time derivatives are proportional to
$\dot\rho$ and $\dot\rho_*$.  Below we will confirm that $\dot\rho = 
-3H(\rho + p)$, where $H \sim \sqrt{\rho}, \sqrt{\rho_*}$ is the Hubble
parameter.  Therefore time derivatives of the perturbations are higher
order in $\rho$ and $\rho_*$ than are $y$ derivatives, and can be neglected
at leading order (except in the (05) Einstein equation, where $\rho^{3/2}$
{\it is} the leading order).  Using
this approximation, we can write the combinations (00), (00)$-$(ii),
(05) and (55) of the Einstein equations as
\beqa
\label{00eq}
	4 A_0' \Psi - \Psi'  &=& 
	\left({\dot a_0\over a_0}\right)^2 b_0^2 e^{2A_0} \\
\label{00iieq}
	- 4 A_0'\Upsilon  + \Upsilon' &=& 
 	2\left( \left({\dot a_0\over a_0}\right)^2 -
 	{\ddot a_0\over a_0}\right) 	b_0^2 e^{2A_0} \\
\label{05eq}
	- {\dot a_0\over a_0}\Upsilon + \dot\Psi  &=& 
	0\\
\label{55eq}
	A_0'(4\Psi + \Upsilon) + \frac{\kappa^2}{3}
        \left( \Phi_0''\delta \Phi-\Phi_0'\delta \Phi'+
        \Phi_0'^{2}\frac{\delta b}{b_{0}} \right)  &=& 
        \left( \left({\dot a_0\over a_0}\right)^2 +
        {\ddot a_0\over a_0}\right)b_0^2 e^{2A_0}
\eeqa
In addition, there is the scalar field equation,
\beq
\label{Phieq}
	\delta\Phi'' = 
	(4\Psi + \Upsilon)\Phi_0'+\left(\frac{4\kappa^2}{3} \Phi_0'^{2}
        + b_0^2 V''(\Phi_0)\right) \delta\Phi + 4A_0'\delta\Phi'
	+\left( 2b_0^{2} V'(\Phi_0)+ 4 A_0' \Phi_0'\right)\frac{\delta b}{b_0}
	 +\Phi_0' {\delta b'\over
b_0}
\eeq	
Assuming $Z_2$ symmetry (all functions symmetric under $y\to-y$),
the boundary conditions implied by the delta function sources at the
branes are
\beqa
\label{Psibc}
\Psi(t,0) &=& +\frac{\kappa^{2}}{6}b_{0}\rho_{*}(t);\qquad\qquad
\qquad\Psi(t,1)=-\frac{\kappa^{2}}{6}b_{0}\rho(t) \\
\label{Upsbc}
 \Upsilon(t,0) &=& -\frac{\kappa^{2}}{2}b_{0}(\rho_{*}+p_*)(t);\quad
\qquad \Upsilon(t,1)= +\frac{\kappa^{2}}{2}b_{0}(\rho+p)(t)\\
\label{Phibc}
 \delta\Phi'(t,y_n) &=& \frac{\delta b(t,y_n)}{b_0}\Phi_0'(t,y_n) +
 (-1)^{n}\left(\frac{b_0}{2}\right)
V_n''(\Phi_0(t,y_n))\,\delta\Phi(t,y_n),
\eeqa
where  in (\ref{Phibc}) $n=0,1$, $y_0 = 0$ and $y_1 = 1$.

{\bf 4. Solutions.} Naively, it would appear that we have five equations
for four unknown perturbations, but of course since gravity is a gauge
theory, this is not the case.  First, we have the relation
${\partial\over\partial t}$[Eq.\ \ref{00eq}] $+ {\dot a_0\over a_0}$[Eq.\
\ref{00iieq}] $=$ [Eq.\ \ref{05eq}].  Furthermore, the (55) Einstein
equation and the scalar equation can be shown to be equivalent, using (00),(ii)
and the zeroth order relations (\ref{zeroeq}): [Eq.\ (\ref{55eq})]$' -
4A_0'\times$[Eq.\ (\ref{55eq}$)]= \Phi_0^{\prime}\times$[Eq.\
(\ref{Phieq})]. So our system is actually underdetermined because of
unfixed gauge degrees of freedom.  To see this more directly, consider an
infinitesimal diffeomorphism which leaves the coordinate positions of the
branes unchanged: $y = \bar y + f(\bar y)$, where $f(0) = f(1) = 0$.  The
metric and scalar perturbations transform as
\beqa
\delta A &\to& \delta A + A_0' f;\qquad \delta N \to \delta N +A_0' f
\nonumber\\
\label{Phishift}
\delta b &\to& \delta b + b_0 f';\qquad \delta\Phi \to \delta\Phi + \Phi_0'
f
\eeqa
If desired, one can form the gauge invariant combinations
\beq
	\delta A' - A'_0 {\delta b\over b_0}-\frac{\kappa^2}{3} \Phi_0' \delta\Phi;
	\qquad	\delta N' - \delta A'; \qquad
	\Phi_0''\delta \Phi-\Phi_0'\delta \Phi'+
        \Phi_0'^{2}\frac{\delta b}{b_{0}}
\eeq
the first two are precisely our variables $\Psi$ and $\Upsilon$ 
and the last one appears in (55) equation.  In terms of these gauge
invariant variables, the system of equations closes.

 It is now easy to verify the following
solution from the (00) and (00)-(ii) equations, {\it i.e.,} eqs.\
(\ref{00eq}-\ref{00iieq}).  
Denoting the warp factor $\Omega = e^{-A_0(1)}$, we find
\beqa
	\Psi &=& {\kappa^2 b_0\over 6(1-\Omega^2)}\, e^{4 A_0(y)}
	\left(
	F(y) (\Omega^4\rho + \rho_*) - 
	(\Omega^4\rho + \Omega^2\rho_*)\right)\\
	\Upsilon &=& {\kappa^2 b_0\over 2(1-\Omega^2)}\, e^{4
	A_0(y)} \left(
	-F(y) (\Omega^4(\rho+p) + \rho_*+p_*)
	+ (\Omega^4(\rho+p) + \Omega^2(\rho_*+p_*))\right)
\eeqa	
where
\beq
\label{Feq}
	F(y) = 1 - (1-\Omega^2) {\int_0^y e^{-2 A_0} dy \over
	\int_0^1 e^{-2 A_0} dy } \cong  e^{-2 k b_0 y}
\eeq
and the Friedmann equations are
\beqa	
\label{F1eq}
	\left({\dot a_0\over a_0}\right)^2 &=& {8\pi G\over 3}\left(\rho_*
	+ \Omega^4 \rho\right)\\
\label{F2eq}
	 \left({\dot a_0\over a_0}\right)^2 -
	 {\ddot a_0\over a_0}  &=& {4\pi G}\left(\rho_* + p_*
        + \Omega^4 ( \rho+p)\right)\\
\label{Geq}
	8\pi G &=& \kappa^2 \left( 2 b_0 \int_0^1 e^{-2 A_0} dy
\right)^{-1} \cong \kappa^2 k  (1-\Omega^{2})^{-1}.
\eeqa
The approximations in eqs.\ (\ref{Feq}) and (\ref{Geq}) hold when the
back reaction of the scalar field on the metric can be neglected.  

In the Friedmann equations (\ref{F1eq}-\ref{F2eq}), we note that $\rho$ is
the bare value of the energy density on the TeV brane, naturally of order
$M_p^4$, while $\Omega^4\rho$ is the physically observable value, of order
(TeV)$^4$.  Since $\rho_*$ has no such suppression, it seems highly
unlikely that $\rho_*$ should be nonzero today; otherwise it would tend to
vastly dominate the present expansion of the universe.  We also point out
that these equations are consistent only if energy is separately conserved
on each brane: $\dot\rho + 3H(\rho+p) = 0$ and $\dot\rho_* +
3H(\rho_*+p_*) = 0$.  This can be derived directly by considering the (05)
Einstein equation, evaluated at either of the branes. The equations of
state on the two branes are completely independent; there is no relation
between $p/\rho$ and $p_*/\rho_*$.\\

{\bf 5. Stiff potential limit.} 
The above solutions are quite general, but they are not complete because
we have not yet solved for the scalar field perturbation, $\delta\Phi$.
This would generically be intractable, but there is a special case 
in which things simplify, namely, when the brane potentials $V_i(\Phi)$
become stiff.  In this case, the boundary condition for the scalar
fluctuation becomes $\delta\Phi = 0$ at either brane.  There is no
information about the derivative $\delta\Phi'$ in this case; although
$\delta\Phi\to 0$, at the same $V''(\Phi)\to\infty$ in such a way that
the product $\delta\Phi V''(\Phi)$ remains finite, and eq.\ (\ref{Phibc})
is automatically satisfied.

Notice that the shift in $\delta\Phi$, eq.\ (\ref{Phishift}), respects the
boundary conditions on
$\delta\Phi$.  Moreover, $\Phi_0'$ is always nonzero for our solution.
It is therefore always possible, given some solution $\delta\Phi$ which 
vanishes at the branes, to
choose an $f$ such that $\delta\Phi$ becomes zero.  This is a convenient
choice of gauge because it simplifies the equations of motion, and we will
make it for the remainder of this letter.\footnote{The above argument is
strictly true only for diffeomorphisms which are constant in time, while
for our problem we need $f(t,y)\sim \rho(t), \rho_*(t)$.  However, the
time variation of such an $f$ is of higher order in $\rho$ and $\rho_*$,
so we can neglect it to leading order in the perturbations.}  
Thus far we have satisfied the (00), (ii) and (05) Einstein equations.  
As noted above, eqs.\ (\ref{55eq}) and (\ref{Phieq}) are equivalent,
so either one just determines the shift in the radius.  Using the former,
and defining 
\beq 
G(y) = \left[\sfrac12
e^{2A_0(y)} + A_0' e^{4A_0(y)} \int_0^y e^{-2A_0} dy \right] / \int_0^1
e^{-2A_0} dy \cong {kb_0 e^{4kb_0y}\over 1-\Omega^2}, 
\eeq 
we find that 
\beqa
	{\delta b\over b_0} &=& {b_0\over 2\Phi_0'^2} \left[
	\Omega^4(\rho-3p) G + (\rho_*-3p_*)(G-A_0'e^{4A_0})\right]
	\nonumber\\
	&\cong&  {k b_0^2 e^{4kb_0y} \over
	2\Phi_0'^2 (1-\Omega^2)}
	\, \left[ \Omega^4(\rho-3p) + \Omega^2(\rho_*-3p_*)\right];
\eeqa
the last expression is found by approximating $A_0 = kb_0y$ everywhere,
which means neglecting the back reaction.  Using the zeroth order solution
(\ref{approxsoln}) for $\Phi_0$, and integrating over $y$,
we can obtain the shift in the size of the extra dimension,
\beq
	\int_0^1 \delta b\, dy \cong {\left[ \Omega^4(\rho-3p) +
	\Omega^2(\rho_*-3p_*)\right] \over 8 (\eps k v_0)^2
	\Omega^{4+2\eps}}
\eeq
We can compare this to the result of ref.\ \cite{CGK} by using their
result for the radion mass, $m^2_r \cong (4/3)\kappa^2(\eps v_0 k)^2
\Omega^{2+2\eps}$, and the relation $k\kappa^2\cong 1/M_p^2$.
Then
\beq
        {\int_0^1 \delta b\, dy\over b_0} \cong {\left[
\Omega^4(\rho-3p) +
       \Omega^2 (\rho_*-3p_*)\right] \over 6 kb_0 m_r^2 M_p^2 \Omega^{2} }
\eeq
which agrees with ref.\ \cite{CGK}, except for small corrections of
order $(1+\Omega^2)$.\footnote{Our correction factor is
$(1-\Omega^{4+2\eps})/(1-\Omega^2)$, while that of ref.\ \cite{CGK}
is $(1-\Omega^2)$.}
As is well known, the shift in the radion
vanishes when the universe is radiation dominated, because the radion 
couples to the trace of the stress energy tensor, which vanishes if
the matter is conformally invariant.

{\bf 6. Implications.}  Above we focused on the shift in the size of the
extra dimension due to cosmological expansion, but the more experimentally
relevant quantity is the shift in the lapse function, $n(t,1)$, evaluated
on the TeV brane.  As emphasized in ref.\ \cite{KOP2}, the change in 
$n(t,1)$ between the present and the past determines how much physical
energy scales on our brane, like the weak scale, $M_W$, have evolved.  The
time dependence of $M_W$ is given by ${M_W(t)/M_W(t_0)} = 
e^{-\delta N(t,1) + \delta N(t_0,1)+\delta N(t,0) - \delta N(t_0,0)}$.
In terms of the variables of the previous section, $\delta N^{\prime} = 
\Psi + \Upsilon + A_0'\delta b/b_0$.    We find that
\beq
	\int_0^1 \delta N^{\prime}(t,y) dy \cong 
	\frac{\kappa^2 b_0}{24kb_0} 
	\left((2\rho+3p) - e^{2kb} (2\rho_*+3p_*) \right)
	+  \int_0^1 A_0'{\delta b\over b_0} dy
\eeq
Interestingly, the new non-$\delta b$ contribution is
present even during radiation domination,
is parametrically smaller than the radion part only in the
matter dominated era, and then only if the back reaction is small
($\epsilon\ll 1$).  If $M_p\Omega\sim 1$ TeV, the shift in
the energy scale since nucleosynthesis is negligible, and this is the only
cosmological constraint on $\delta N$.  However, near the weak scale,
the effect could be more interesting.  To first order in the physical
energy density $\rho_p = \Omega^4\rho$, at early times,
\beq
	M_W(t) \cong M_W(t_0)\left(1 - {\rho_p(t)\over 8 \Omega^4
	M_p^2 k^2}\right),
\eeq
assuming that $\rho_*=p_*=0$.
It is conceivable that $k \sim M_p/30$---in fact, the RS model requires $k<
M_P$ for consistency, so that higher dimension operators in the
gravitational part of the action do not become important. 
With these parameters and assuming $g_*\sim 100$ relativistic
degrees of freedom and $\Omega M_p = 1$ TeV, the correction to $M_W$
becomes of order unity at a temperature of 130 GeV.  Thus the temporal
variation in fundamental mass scales might have some relevance for the
electroweak phase transition and baryogenesis.

\bigskip

We thank C.\ Csaki, M.\ Graesser and G.\ Kribs for helpful discussions.
JC thanks Nordita for its hospitality while this work was being finished.


\end{document}